\documentclass[usenatbib,letter]{aa}

\usepackage{graphicx}
\usepackage{txfonts}
\usepackage{hyperref}
\usepackage{xcolor}
\hypersetup{
colorlinks,
linkcolor={red!80!black},
citecolor={blue!80!black},
urlcolor={blue!80!black}
}

\usepackage{physics}
\usepackage{subcaption}
\usepackage{soul}

\newenvironment{myfont}{\fontfamily{ppl}\selectfont}{\par}
\DeclareTextFontCommand{\textmyfont}{\myfont}

\newcommand{\code}[1]{\texttt{#1}}

\def\nifs{\iso{56}Ni}

\def\cm3{cm$^{-3}$}
\def\kms{\mbox{km~s$^{-1}$}}

\def\msun{$M_{\odot}$}

\def\one{\ts {\,\sc i}}
\def\two{\ts {\,\sc ii}}
\def\three{\ts {\,\sc iii}}

\def\beq{\begin{equation}}
\def\eeq{\end{equation}}

\def\lesssim{\mathrel{\hbox{\rlap{\hbox{\lower4pt\hbox{$\sim$}}}\hbox{$<$}}}}
\def\gtrsim{\mathrel{\hbox{\rlap{\hbox{\lower4pt\hbox{$\sim$}}}\hbox{$>$}}}}

\def\one{{\,\sc i}}
\def\two{{\,\sc ii}}
\def\three{{\,\sc iii}}

\def\v1d{{\code{V1D}}}

\def\cmfgen{{\code{CMFGEN}}}

\def\ergs{erg\,s$^{-1}$}

\def\mic{\,$\mu$m}

\def\sn{SN\,2024ggi}

\def\oidoub{[O\one]\,0.632\,$\mu$m}

\def\neiifs{[Ne\two]\,12.810\,$\mu$m}
\def\neiiifs{[Ne\three]\,15.550\,$\mu$m}

\def\ariimir{[Ar\two]\,6.983\,$\mu$m}

\def\caiidoub{[Ca\two]\,0.731\,$\mu$m}

\def\nkiiopt{[Ni\two]\,0.738\,$\mu$m}
\def\nkiinir{[Ni\two]\,1.939\,$\mu$m}
\def\nkiimir{[Ni\two]\,6.634\,$\mu$m}

\newcommand{\iso}[2]{\ensuremath{^{#1}\rm{#2}}}

\begin{document}

   \title{An optical to infrared study of type II SN\,2024ggi at nebular times}

   \titlerunning{Optical-to-infrared study of \sn\ at nebular times}

   \author{
     Luc Dessart\inst{\ref{inst1}}
     \and
     Rubina Kotak\inst{\ref{inst2}}
     \and
     Wynn Jacobson-Gal\'an\inst{\ref{inst3},\ref{inst4}}
     \and
     Kaustav Das\inst{\ref{inst3}}
     \and
     Christoffer Fremling\inst{\ref{inst5},\ref{inst3}}   
     \and
     Mansi Kasliwal\inst{\ref{inst3}}
     \and
     Yu-Jing Qin\inst{\ref{inst3}}
     \and
     Sam Rose\inst{\ref{inst3}}   
  }

\institute{
  Institut d'Astrophysique de Paris, CNRS-Sorbonne Universit\'e, 98 bis boulevard Arago, F-75014 Paris, France\label{inst1}
  \and
  Department of Physics and Astronomy, FI-20014, University of Turku, Finland\label{inst2}
  \and
  Cahill Center for Astrophysics, California Institute of Technology, Pasadena, CA 91125, USA\label{inst3}
  \and
  NASA Hubble Fellow\label{inst4}
  \and
  Caltech Optical Observatories, California Institute of Technology, Pasadena, CA 91125, USA\label{inst5}
  }

   \date{}

  \abstract{
We present 0.3--21\mic\ observations at $\sim$\,275\,d and $\sim$\,400\,d for Type II supernova (SN) 2024ggi, combining ground-based optical and near-infrared data from the Keck I/II telescopes and space-based infrared data from the James Webb Space Telescope. Although the optical regions dominate the observed flux, \sn\ is bright at infrared wavelengths (65\,\%/35\,\% falls each side of 1\mic). \sn\ exhibits a plethora of emission lines from H, He, intermediate-mass elements (O, Na, Mg, S, Ar, Ca), and iron-group elements (IGEs; Fe, Co, and Ni) -- all lines have essentially the same width, suggesting efficient macroscopic chemical mixing of the inner ejecta at $\lesssim$\,2000\,\kms\ and little mixing of \nifs\ at larger velocities. Molecular emission in the infrared range is dominated by the CO fundamental, which radiates about 5\,\% of the total SN luminosity. A molecule-free radiative-transfer model based on a standard red-supergiant star explosion (i.e., $\sim$\,$10^{51}$\,erg, 0.06\,\msun\ of \nifs\ from a 15.2\,\msun\ progenitor) yields a satisfactory match throughout the optical and infrared at both epochs. The \sn\ CO luminosity is comparable to the fractional decay-power absorbed in the model C/O-rich shell -- accounting for CO cooling would likely resolve the model overestimate of the \oidoub\ flux. The relative weakness of the molecular emission in \sn\ and the good overall match obtained with our molecule-free model suggests negligible microscopic mixing -- about 95\,\% of the SN luminosity is radiated by atoms and ions. Lines from IGEs, which form from explosion ashes at such late times, are ideal diagnostics of the magnitude of \nifs\ mixing in core-collapse SN ejecta. Stable Ni, clearly identified in \sn\ (e.g., \nkiimir), is probably a common product of massive-star explosions.
}

   \maketitle

\section{Introduction}

Nebular-phase, panchromatic observations of the low-energy radiation from core-collapse supernovae (SNe) are a major step  towards a complete characterization of the ejecta composition, which carries a wealth of information on the preSN evolution and explosive nucleosynthesis. So far, such a comprehensive spectroscopic dataset extending into the infrared has been limited to SN\,1987A, observed by the Kuiper Airborne Observatory \citep{wooden_87A_ir_93}, and a few Type II-Plateau SNe observed by the Spitzer Space Telescope (e.g., \citealt{kotak_04dj_05,kotak_05af_06,kotak_04et_09}; \citealt{fabbri_dust_11}; \citealt{meikle_04dj_11}). Such optical to infrared observations contain essentially 100\,\% of the SN radiation in noninteracting SNe II, capture the weak  emission from underabundant metals such as Ar or Ni typically unavailable in the optical, and allow for a complete census of atoms, ions, and molecules cooling the decay-powered ejecta at nebular times \citep{jerkstrand_04et_12,liljegren_co_20,dessart_sn2p_21,liljegren_ibc_mol_23,mcleod_mol_24,dessart_ir_25}.

In this letter, we report on the nearby Type II \sn\ for which we obtained Keck/LRIS optical, Keck/NIRES near-infrared, and James Webb Space Telescope (JWST) NIRSPEC/MIRI infrared observations at two nebular epochs of about 300 and 400\,d after explosion (see Section~\ref{sect_obs} for a complete description of the observational dataset). This object received much attention because it was discovered as the shock broke out from the stellar surface \citep{wynn_24ggi_24,shrestha_24ggi_24,zhang_24ggi_24,chen_24ggi_early_24,chen_24ggi_25}. This breakout persisted over days as the shock crossed an extended, dense circumstellar material, as evidenced by the presence of narrow emission lines, whose widths grew due the acceleration imparted by the SN radiation \citep{pessi_24ggi_24,shrestha_24ggi_24,dessart_radacc_25}. After 1--2 weeks, \sn\ evolved into a standard Type II SN \citep{ertini_24ggi_25}, with properties reminiscent of SN\,1999em \citep{leonard_99em,elmhamdi_99em_03,d13_sn2p}.

In the next section, we analyze our observational dataset of \sn\ and confront it to the radiative-transfer models of \citet{dessart_ir_25}. We then present our conclusions in Section~\ref{sect_conc}. Throughout this work, we adopted the following observational characteristics for \sn: a reddening $E(B-V)$ of 0.07\,mag due to the Milky Way \citep{schlegel_dust_mw_98,schlafly_finkbeiner_11}  and 0.084\,mag due to the host galaxy \citep{wynn_24ggi_24}, a redshift of 0.002215 \citep{wynn_24ggi_24}, a redshift-independent host-galaxy distance of 7.24\,Mpc \citep{saha_dist_06}, and a time of first-light at MJD\,60410.56 \citep{wynn_24ggi_24}. Supplementary information on this work is provided in the appendix.

\begin{figure*}
\centering
\includegraphics[width=0.8\hsize]{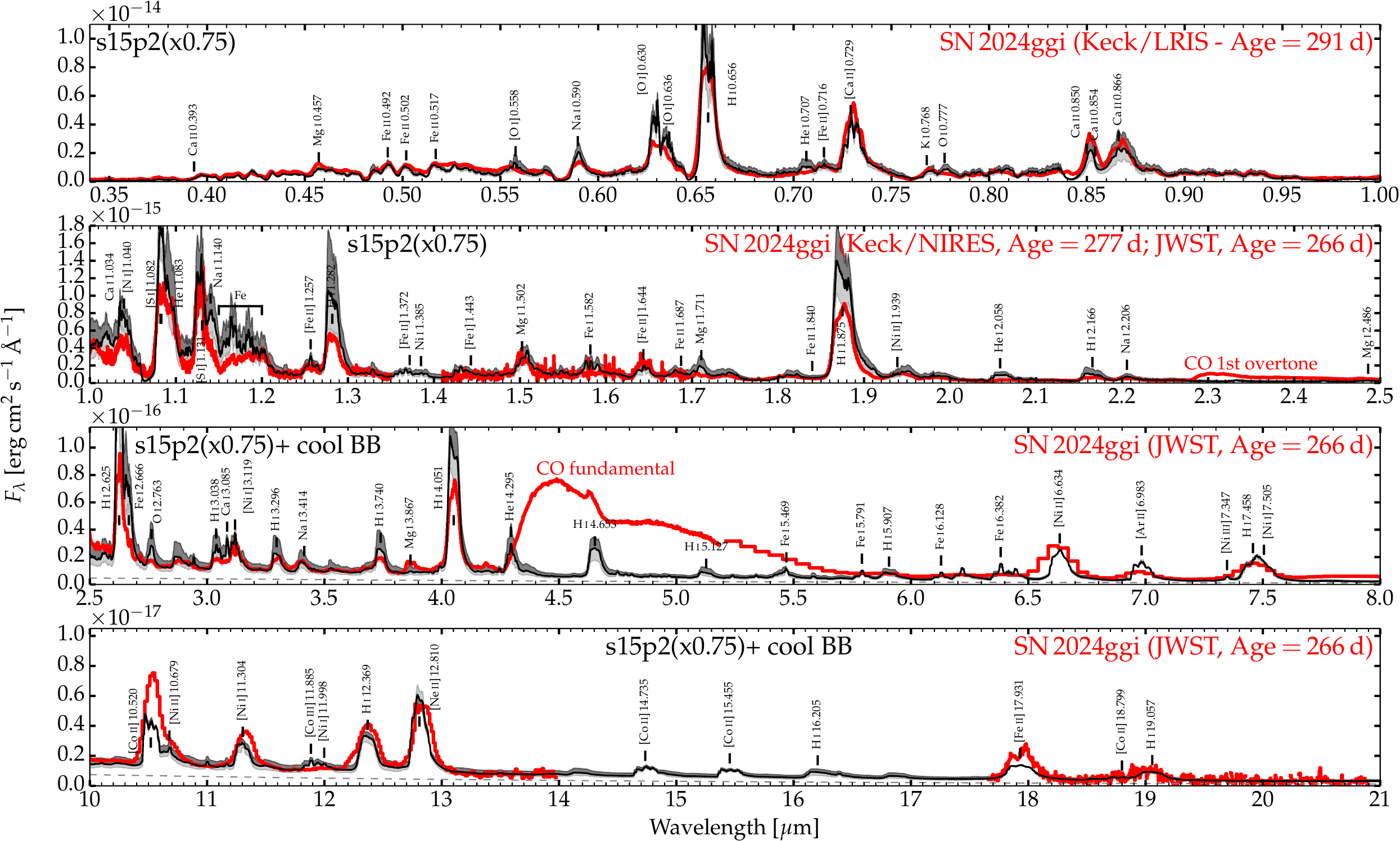}
\caption{Comparison between the optical to infrared spectroscopic observations of \sn\ at 266--291\,d (red; see Section~\ref{sect_obs}) with model s15p2 interpolated to the same epoch and scaled by a factor of 0.75 (black), each panel covering from top to bottom the optical, near-infrared and mid-infrared ranges. The data were corrected for redshift and reddening and the model was scaled to the SN distance. The flux shift corresponding to the same model but at $\pm$\,30\,d is shown as light/dark grey. Labels indicate the main emission features. For easier comparison, the radiative-transfer model flux was augmented by a composite blackbody spectrum, which may arise from molecular emission (dashed; the blackbody temperatures and radii are 1230/400\,K and $0.8/5.0\times 10^{15}$\,cm). The region between 8 and 10\mic, which appears featureless with only weak SiO emission in \sn, is shown in Fig.~\ref{fig_obs_only}.
\label{fig_mod_vs_obs_jan25}
}
\end{figure*}

\begin{figure*}
\centering
\includegraphics[width=0.8\hsize]{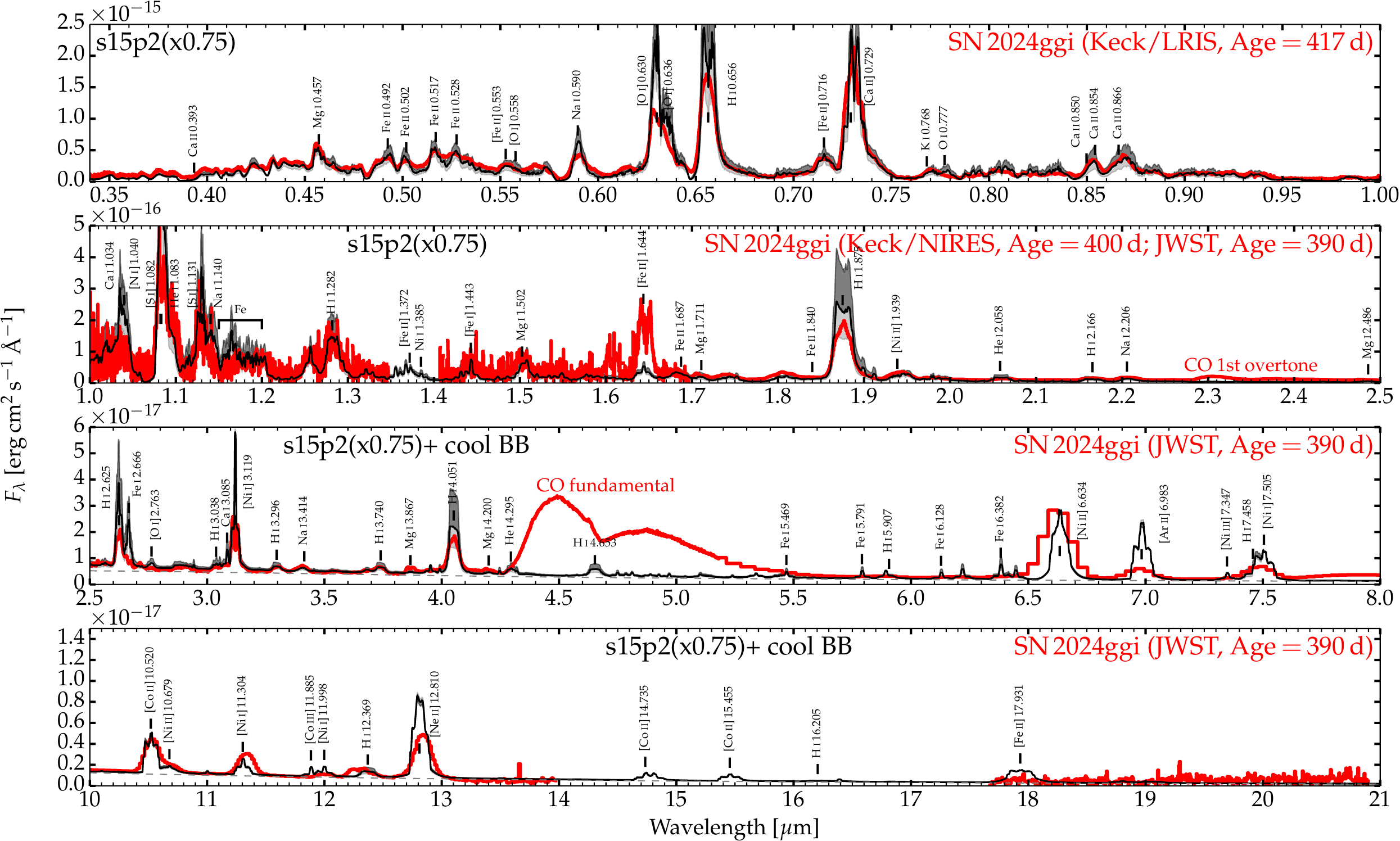}
\caption{Same as Fig.~\ref{fig_mod_vs_obs_jan25} but now for the observations of \sn\ at around 400\,d and the contemporaneous model s15p2.
\label{fig_mod_vs_obs_may25}
}
\end{figure*}

\section{Analysis}
\label{sect_analysis}

Figures~\ref{fig_mod_vs_obs_jan25} and \ref{fig_mod_vs_obs_may25} present the full optical to infrared observations of \sn\ at $\sim$\,275\,d and $\sim$\,400\,d, covering continuously from 0.32 up to $\sim$\,21\mic\ apart from a few gaps due to atmospheric absorption (NIRES data) or our selected instrumental setup (JWST data). These data are compared to the s15p2 radiative-transfer model from \citet{dessart_ir_25}, computed with \cmfgen\ \citep{HD12}, originally presented in \citet{dessart_sn2p_21}, and based on the s15.2 explosion model of \citet{sukhbold_ccsn_16}. This model is for a 15.2\,\msun\ star evolved at solar metallicity that reached core collapse as a 12.58\,\msun\ red-supergiant star and following explosion ejected 10.95\,\msun\ of material with $0.84 \times 10^{51}$\,erg and 0.063\,\msun\ of \nifs. These properties are comparable to those inferred by \citet{ertini_24ggi_25} from radiation-hydrodynamics modeling of the light curve and the photospheric velocity evolution. Our radiative-transfer model only treats atoms and ions in the ejecta material, and thus ignores molecules and dust. The lack of a peculiar, blue optical color or broad, boxy emission profiles \citep{dessart_csm_22,dessart_late_23} suggests that there is little interaction with circumstellar material in \sn\ at these epochs (see, e.g., SN\,2023ixf where such signatures are identified; \citealt{folatelli_23ixf_25,wynn_iii_25,kumar_23ixf_25,zheng_23ixf_25}). We thus accounted only for radioactive decay power in our s15p2 model (specifically, we accounted for the two-step decay chains whose parent isotopes are \iso{44}Ti, \iso{48}Cr, \iso{52}Fe, \nifs, and \iso{57}Ni; see discussion in \citealt{dessart_ir_25}).

These nebular spectra of \sn\ exhibit H\one\ lines from the Balmer (i.e., H$\alpha$ and H$\beta$), through to the Paschen (e.g., 1.875 and 1.282\mic), Brackett (e.g., 4.051, 2.625, and 2.166\mic), Pfund (7.458, 4.653, and 3.740\mic), Humphreys (e.g., 12.369\mic), and up to the seventh series with 19.057\mic. The strong He\one\,1.083\mic\ contributes to the emission at 1.08\mic\ (another contributor is [S\one]\,1.082\mic), but other He\one\ lines are weak (e.g., at 0.707\mic) or not clearly observed (i.e., at 2.058\mic\ and 4.295\mic). Lines of C and N are essentially invisible, but O contributes a strong optical line with \oidoub\ ([O\one]\,0.558\mic\ and O\one\,0.774\mic\ are weak) with perhaps a weak emission at 2.763\mic. Neon contributes a strong forbidden-line emission at 12.810\mic\ -- \neiiifs\ was not observed in our setup. A few weak Na\one\ lines are present at 1.140, 2.206, and 3.414\mic. Weak Mg\one\ lines are present at 1.502 and 1.711\mic. Sulfur contributes one emission at 1.082\mic\ overlapping with He\one\,1.083\mic. Argon contributes one isolated, forbidden-line emission with [Ar\two]\,6.983\mic. The resonant transition of K\one\ at 0.768\mic\ is also present (not to be confused with O\one\ at 0.777\mic). Calcium contributes mostly in the optical with \caiidoub\ and the Ca\two\ near-infrared triplet. There are numerous, typically isolated emission lines from Fe, Co, and Ni in the infared, in contrast with the optical where iron-group elements contribute primarily with forests of lines of Fe\two\ and Fe\one\ -- one exception is \nkiiopt\ although it is made ambiguous because of its overlap with \caiidoub\ as well as a background of Fe lines. We identify a forest of Fe\one\ and Fe\two\ lines up to two microns, [Fe\two] at 17.931\mic, Co with [Co\two]\,10.520\mic, and a number of Ni lines with [Ni\two] at 1.939, 6.634, and 10.679\mic, and [Ni\one] at 3.119, 7.505 and 11.304\mic.

\begin{figure}
\centering
\includegraphics[width=0.9\hsize]{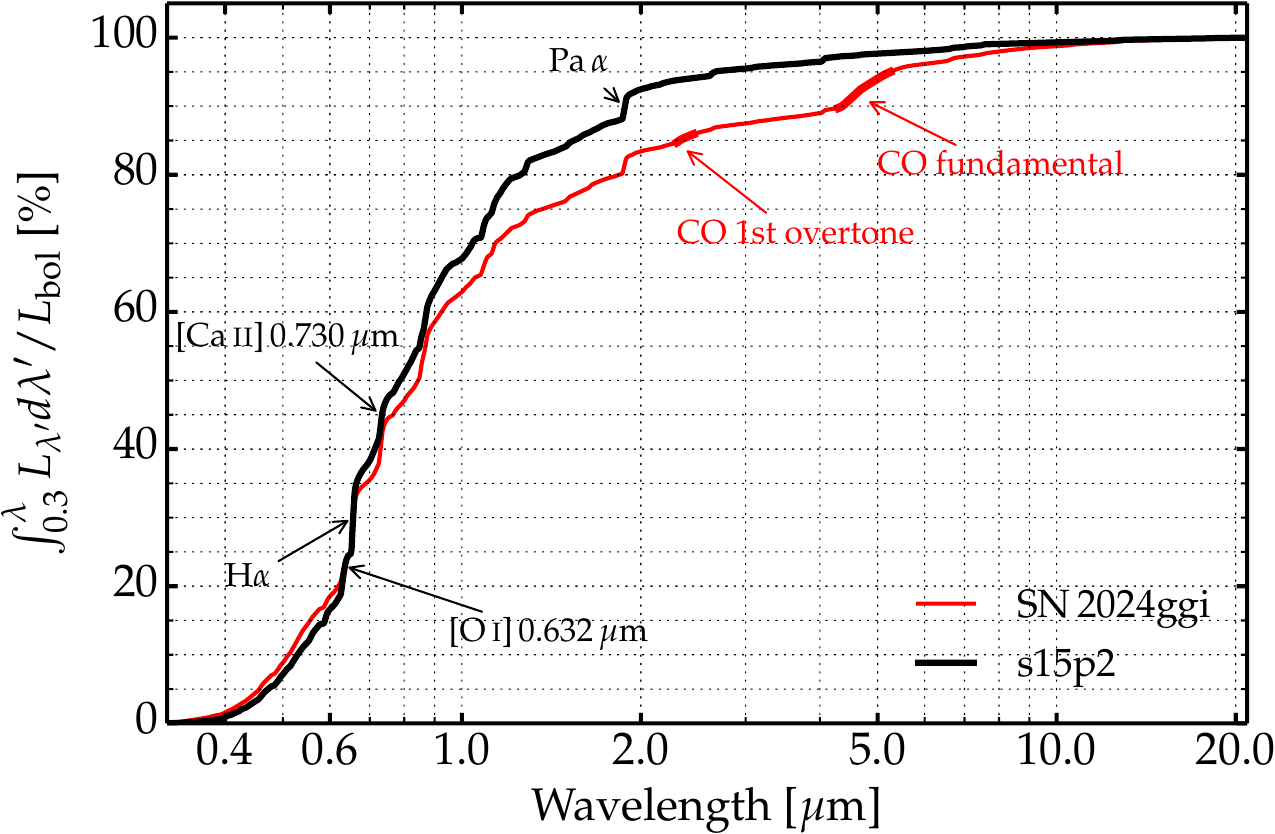}
\caption{Fractional luminosity integrated from 0.3 up to 21\mic\ for the first epoch of observations of \sn\ and model s15p2 at 275\,d. The data were corrected for redshift and reddening. Linear interpolation is used to infer the flux in regions without data. \label{fig_frac_flux_jan25}
}
\end{figure}

The model s15p2 at $\sim$\,275\,d from \citet{dessart_ir_25}, distance-scaled to 7.24\,Mpc (with an additional scaling of 0.75 to match the inferred bolometric luminosity of $5.67 \times 10^{40}$\,\ergs\ of \sn\ -- the original s15p2 model has slightly too much \nifs\ or possibly traps too efficiently the associated $\gamma$ rays) yields a satisfactory match to the observations of \sn\ from the optical through the infrared (Fig.~\ref{fig_mod_vs_obs_jan25}; to give a sense of the temporal evolution, the flux shift from the model predictions 30\,d prior or after are indicated with a dark/light grey shading) -- a similar match is obtained for the second epoch of observations at about 400\,d (Fig.~\ref{fig_mod_vs_obs_may25}; see also Fig. \ref{fig_obs_only} where both observations are shown). This suggests that the overall characteristics of the ejecta (i.e., primarily the yields) are broadly consistent. However, the model is too bright in all H\one\ lines by several 10\,\%, suggesting that too much decay power is absorbed by H-rich material. In our shuffled-shell model, 2.2\,\msun\ from the H-rich ejecta was mixed in, which may be a little large. Lowering this amount to 1.4\,\msun\ reduces in part this discrepancy, by about 10\,\%. Invoking a lower \nifs\ mass might help, although this tends to primarily induce a global reduction of the model flux. A more promising alternative is an enhanced $\gamma$-ray escape, as would result from a lighter progenitor H-rich envelope or a higher ejecta kinetic energy. Weak helium lines are also overestimated. There is a good match to all lines from intermediate mass elements, although the model overestimates \oidoub\ (line luminosity of 3.1 compared to $2.0 \times 10^{39}$\,\ergs) and \neiifs. A lower mass progenitor of 13 or 14\,\msun\ could solve both discrepancies because of their smaller O and Ne yields \citep[see also Section~\ref{sect_add_line_evol} and Fig.~\ref{fig_ar2_ne2_evol}]{dessart_sn2p_21}. However, had we treated molecular cooling in our model, this \oidoub\ emission would be weaker so this overestimate is in practice a desirable offset. CO molecules cannot  form in the C-poor O/Ne/Mg shell (see also \citealt{liljegren_ibc_mol_23}) where \neiifs\ forms so a lower Ne abundance is likely required. The strength of \ariimir\ is overestimated. This line forms in the Fe-rich and Si-rich regions, and to a lesser extent in the He-rich shell, but is a secondary coolant (it contains $\lesssim$\,0.2\,\% of the model luminosity). A lower Ar abundance may resolve this discrepancy -- there is significant heterogeneity in the strength of that line in the sample of core-collapse SNe of \citet{dessart_ir_25}, with model s15p2 yielding the greatest strength (see Fig.~\ref{fig_ar2_ne2_evol} and further discussion in Section~\ref{sect_add_line_evol}). The forest of Fe line emission in the optical as well as the more isolated Fe, Co, and Ni lines in the infrared are well matched by the model. These individual lines have similar shapes, form in regions with $\lesssim$\,2000\kms\ in common with all other elements from H, He and those of intermediate mass, which suggests a complete macroscopic mixing of the preSN shell-like composition structure in the inner ejecta (see Figs~\ref{fig_mod_vs_obs_indiv_lines_epoch1}--\ref{fig_mod_vs_obs_indiv_lines_epoch2}). It confirms that the presence of stable Ni may be rather common in core-collapse SN ejecta -- other examples include SN\,1987A or SN\,2004et, something that is hard to infer from optical spectra with the contaminated \nkiiopt\ line \citep{jerkstrand_ni_15}. These spectral properties thus show how strikingly different are the atomic properties of different species, both in terms of the number of lines and their distribution with wavelength. A vivid representation is given in Fig.~\ref{fig_species_s15p2_at_300d}.

When evaluating the quality of the `fits', one must recall first that the majority of the flux is emitted below 1\mic\ (62\,\% for \sn\ and 68\,\% for the model, both at $\sim$\,275\,d) so optical emission lines, rather than molecules, perform the bulk of the ejecta cooling (Fig.~\ref{fig_frac_flux_jan25}). When raised to below 2\mic, this percentage grows to 84\,\% and 93\,\%. The difference in flux distribution between \sn\ and the s15p2 model arises primarily from CO molecular cooling and in particular the CO fundamental, whose luminosity at $\sim$\,275\,d is $\sim$\,$3 \times 10^{39}$\,\ergs\ (or $\sim$\,5\,\% of the total luminosity) and zero in the model. Interestingly, this corresponds to the fractional decay-power absorbed in the C/O-rich material of our model s15p2 (where CO would form), which would suggest that, had we treated molecules in our calculation, CO would probably have been the exclusive coolant of the C/O-rich material (expansion cooling is subdominant at such late epochs). However, because the O-rich material absorbs $\sim$\,20\,\% of the total decay power, it would only have reduced the model \oidoub\ line strength by only 20--30\,\%, thereby resolving the discrepancy (note that \oidoub\ is not the only coolant of the O-rich material in the model; see \citealt{dessart_sn2p_21}).

\section{Conclusions}
\label{sect_conc}

In this letter, we have presented ground- and space-based, optical to infrared observations at $\sim$\,275\,d and $\sim$\,400\,d of \sn\ together with the s15p2 radiative-transfer model from \citet{dessart_ir_25} based on a model from \citet{sukhbold_ccsn_16} for a standard explosion of a star of 15.2\,\msun\ initially. Whereas the optical range is dominated by a forest of overlapping Fe emission lines and a few strong lines of H, O, and Ca, the infrared reveals a wealth of lines, of comparable width, from H, He, to O, Na, Mg, S, Ar, and Fe, Co, or Ni. This suggests efficient macroscopic chemical mixing of the inner ejecta and little mixing of \nifs\ at larger velocities. Strong CO emission is observed in \sn, corresponding to 5\,\% of the total luminosity. The model fares well against these panchromatic observations despite the neglect of molecules, which supports the notion that CO forms in a very restricted volume of the ejecta (i.e., the C/O-rich material) -- a more uniform distribution of CO would alter the temperature, ionization, and coolants throughout and strongly impact the resulting spectrum (see example in \citealt{mcleod_mol_24}). Assuming that 20--30\,\% of the model \oidoub\ flux should have gone into CO emission, a progenitor mass of $\sim$\,15\,\msun\ seems compatible with the \sn\ spectra. This is in line with the inferences of \citet[14\,\msun]{hueichapan_24ggi_25} and \citet[12--15\,\msun]{ferrari_24ggi_25}, which were based on individual optical line diagnostics or radiative-transfer models that did not include molecular cooling. This value is also close to the inference of \citet[13\,\msun]{xiang_24ggi_24} from pre-explosion imaging photometry, but in tension with the 10.2\,\msun\ from a study of the SN environment \citep{hong_24ggi_24}.

\begin{figure}
\centering
\includegraphics[width=0.9\hsize]{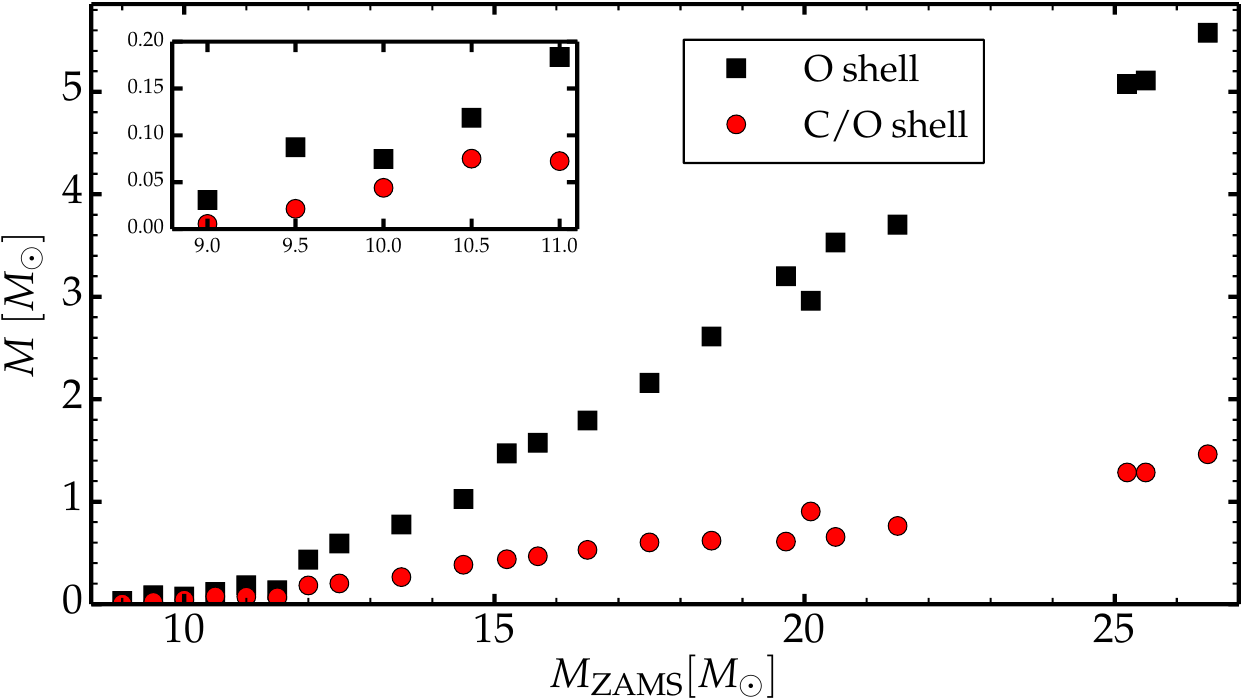}
\caption{Illustration of the mass of the O-rich shell (black) and the C/O-rich shell (red) in the explosion models of \citet{sukhbold_ccsn_16} having a zero-age main sequence mass ($M_{\rm ZAMS}$) in the range 9.0 to 26.5\,\msun. The inset zooms in on the lower mass progenitors and their very low metal yields.
\label{fig_frac_co_sh}
}
\end{figure}

One can evaluate the impact of neglecting CO molecular cooling in core-collapse SN models, in particular on the strength of \oidoub. The maximum CO cooling power possible is the decay-power absorbed in the C/O-rich layers of the ejecta. Assuming \oidoub\ forms throughout the O-rich material and a uniform deposition of $\gamma$-rays in this region, the reduction in \oidoub\ line flux is equal to just $M$(C/O-sh)/$M$(O-sh). As shown in Fig.~\ref{fig_frac_co_sh} for a subset of explosion models from \citet{sukhbold_ccsn_16}, the C/O-rich shell is typically 30\,\% of the O-rich shell in $>$\,12\,\msun\ progenitors, which sets a strict limit of 30\,\% to the reduction of the \oidoub\ flux by CO molecular cooling. In lighter progenitors, because of the low O-shell mass and potentially greater fractional mass occupied by the C/O-rich shell, CO molecular cooling may lead to a severe reduction of the inherently weak \oidoub\ flux, and may explain the existence of SNe II without \oidoub.

\begin{acknowledgements}

  This work is based in part on observations made with the NASA/ESA/CSA James Webb Space Telescope. The data were obtained from the Mikulski Archive for Space Telescopes at the Space Telescope Science Institute, which is operated by the Association of Universities for Research in Astronomy, Inc., under NASA contract NAS 5-03127 for JWST. We are grateful for the allocation of JWST DDT time to programmes 6678 (PI: Kotak) and 6716 (PI: Ashall). Some of the data presented herein were obtained at the W. M. Keck Observatory, which is operated as a scientific partnership among the California Institute of Technology, the University of California, and NASA. The Observatory was made possible by the generous financial support of the W. M. Keck Foundation. The authors wish to recognize and acknowledge the very significant cultural role and reverence that the summit of Maunakea has always had within the indigenous Hawaiian community. We are most fortunate to have the opportunity to conduct observations from this mountain. RK acknowledges support from the Research Council of Finland (340613). W.J.-G. is supported by NASA through Hubble Fellowship grant HSTHF2-51558.001-A awarded by the Space Telescope Science Institute, which is operated for NASA by the Association of Universities for Research in Astronomy, Inc., under contract NAS5-26555. 

\end{acknowledgements}


\onecolumn

\appendix
\label{sect_appendix}

\section{Additional results}
\label{sect_add}

\subsection{Strength of \ariimir, \neiifs, and \oidoub\ in the core-collapse SN models}
\label{sect_add_line_evol}

\begin{figure*}[b]
\centering
\includegraphics[width=0.33\hsize]{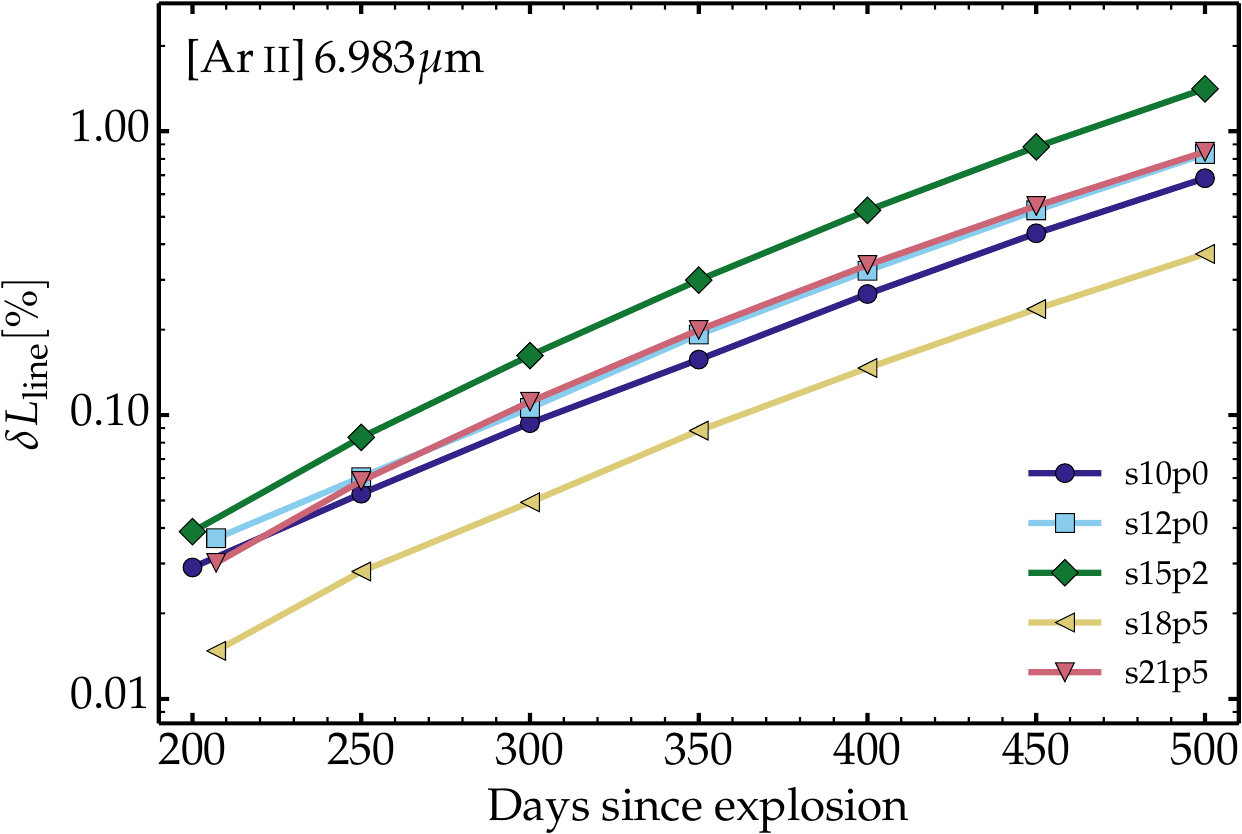}
\includegraphics[width=0.33\hsize]{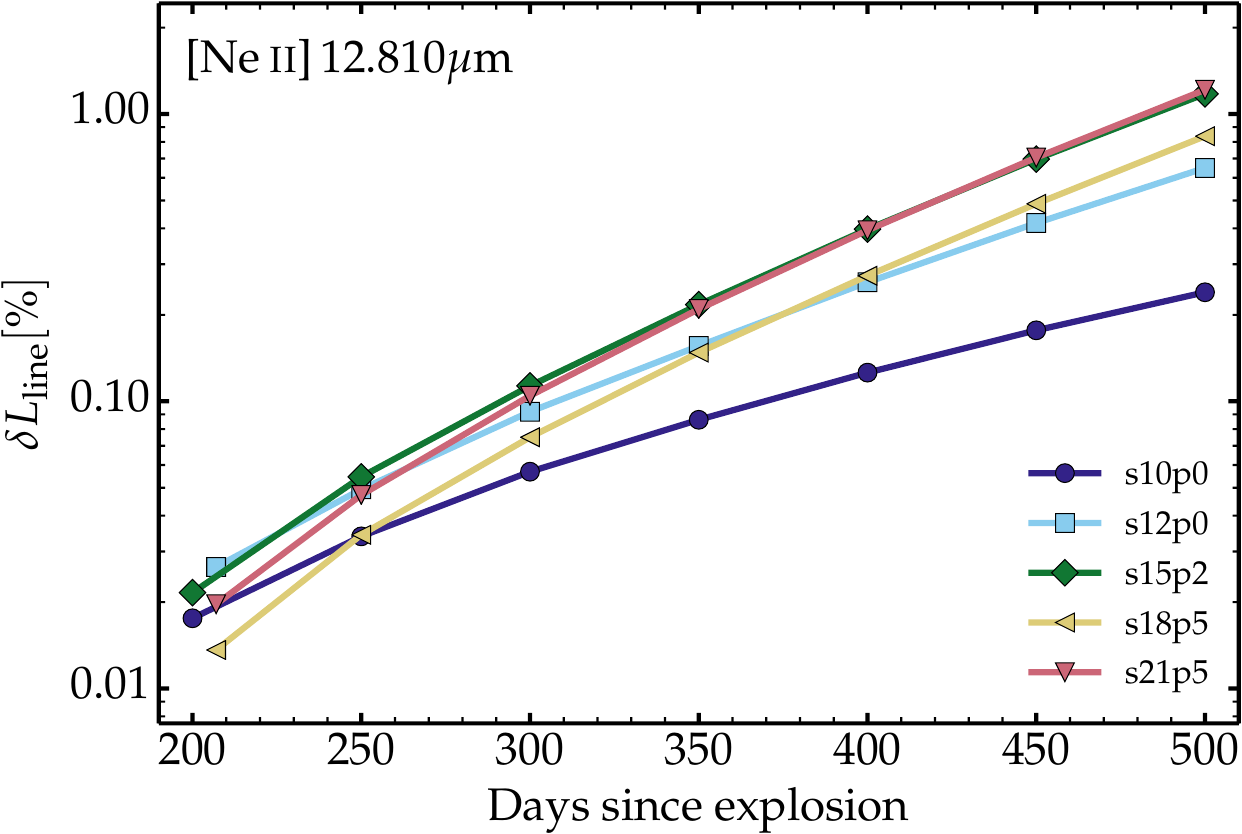}
\includegraphics[width=0.33\hsize]{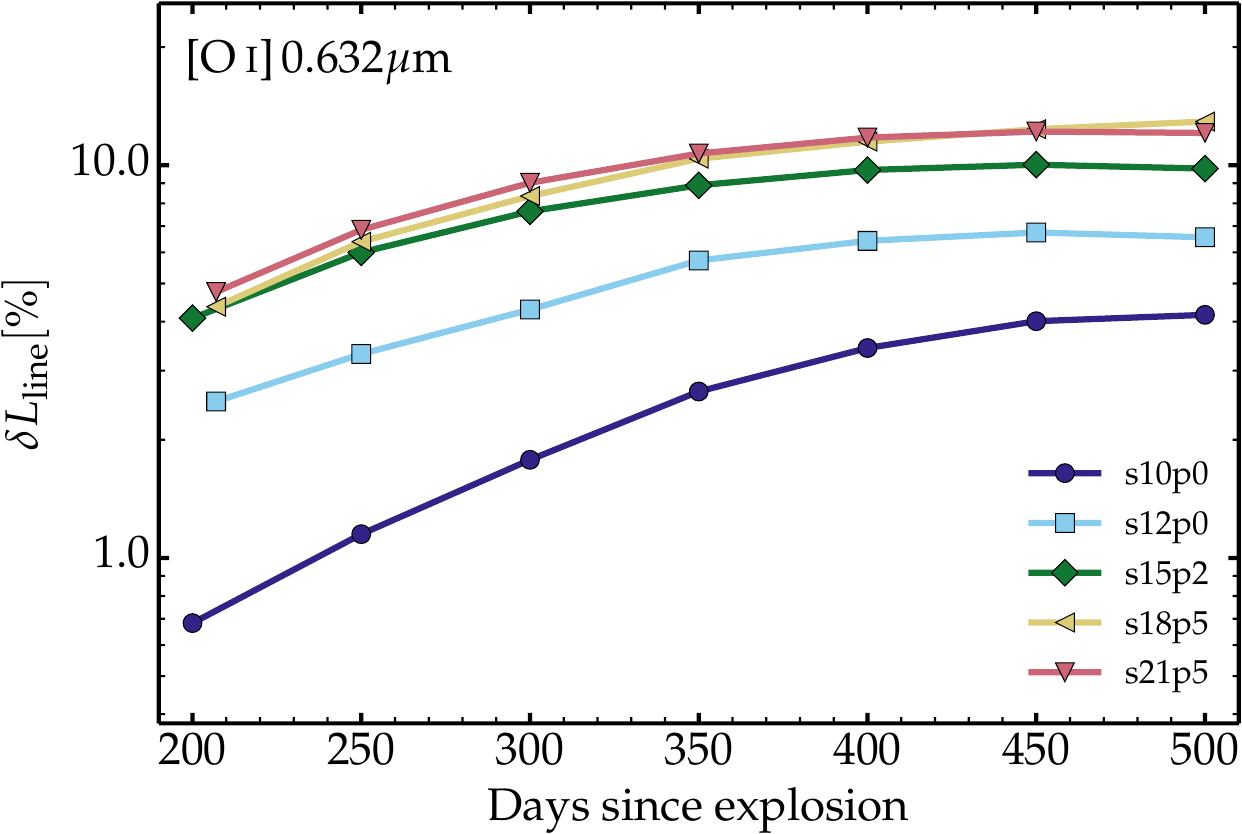}
\caption{Evolution of the percentage fraction of the bolometric flux that is radiated in \ariimir\ (left), \neiifs\ (middle), and \oidoub\ (right) for a sample of core-collapse SN models from \citet{dessart_ir_25}.
\label{fig_ar2_ne2_evol}
}
\end{figure*}

The left panel of Fig.~\ref{fig_ar2_ne2_evol} illustrates the fractional luminosity radiated in \ariimir\ from 200 to 500\,d after explosion in the sample of models presented in \citet{dessart_ir_25}. Line \ariimir\ is strongest (weakest) in model s15p2 (s18p5), partly as a result of the difference in Ar abundance (0.011 vs 0.004\,\msun). There is a lot of scatter in that Ar yield since model s21p5 has 0.016\,\msun\ of Ar whereas models s10p0 and s12p0 are down at about 0.002\,\msun. Variations in Ar abundance, in ionization (driven in part from changes in density and expansion rate), decay-power absorbed, and other coolants (e.g., the Ca abundance also shows a monotonic trend with initial mass, while it is the strongest coolant of the Fe-rich and Si-rich material) may all impact the \ariimir\ line strength.

The middle panel of Fig.~\ref{fig_ar2_ne2_evol} is a counterpart of the left panel for \neiifs. As for \ariimir, model s15p2 shows the strongest \neiifs\ luminosity. Unlike for Ar, the Ne abundance evolves monotonically with initial mass, increasing from $\sim$\,0.01 up to $\sim$\,0.6\,\msun\ for initial masses between 10.0 and 21.5\,\msun. Here, the variation in the strength of \neiifs\ is both driven by this abundance variation and changes in ionization (see Section~5 of \citealt{dessart_ir_25}). So, a slightly lower initial mass and greater density (as would result for example from clumping) could bring the model \neiifs\ into agreement with the observed line in \sn. The right panel of Fig.~\ref{fig_ar2_ne2_evol} shows the fractional luminosity radiated in \oidoub\ for the same sample, which exhibits a monotonic trend with initial mass as well as a strength that is typically ten times greater than \neiifs. These models ignore  cooling by molecules.

\subsection{Elemental contribution to the s15p2 model flux at 300\,d}

\begin{figure*}
\centering
\includegraphics[width=\hsize]{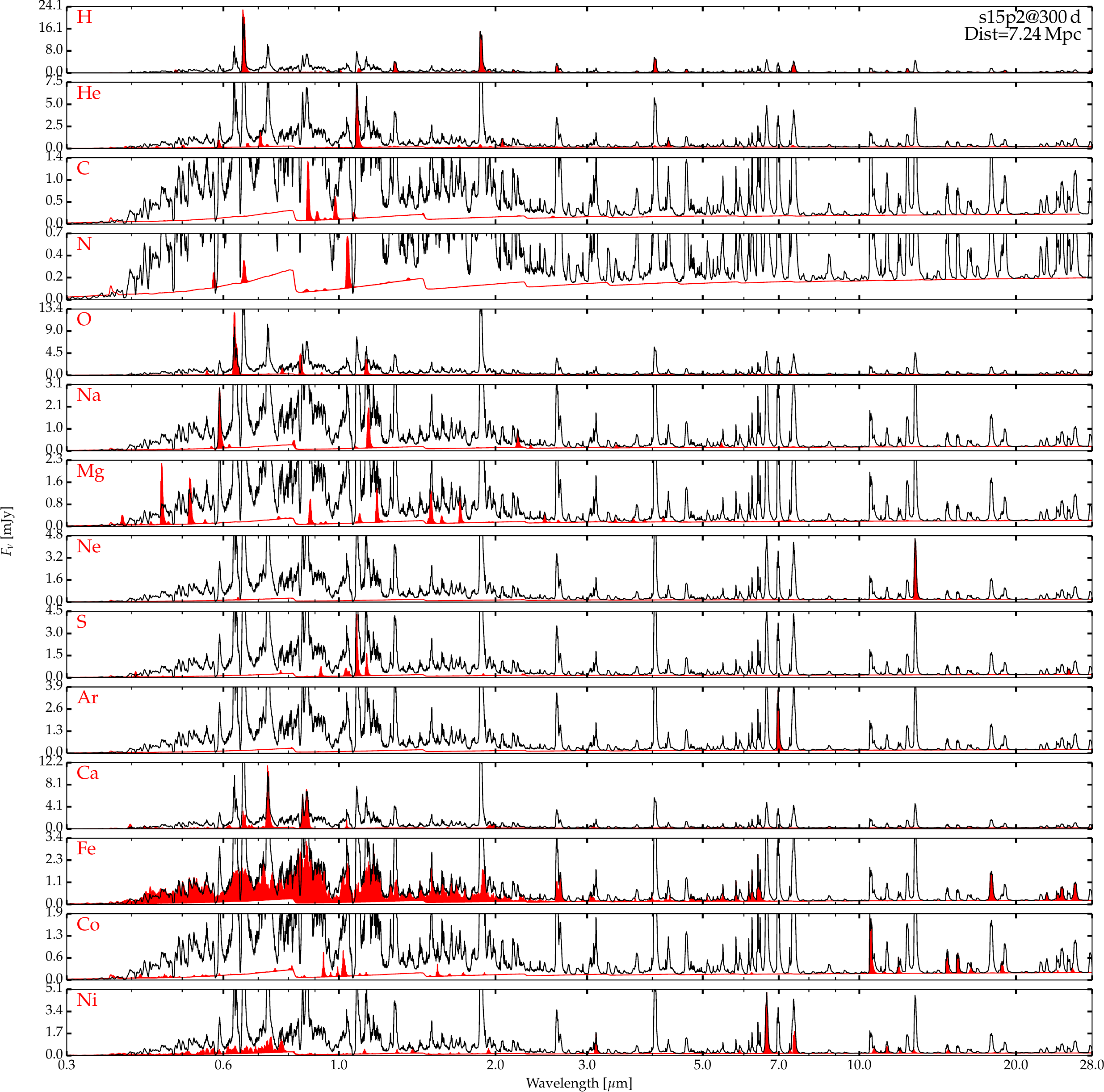}
\caption{Breakup of species contributions to the total flux in model s15p2 at 300\,d. We show a stack of synthetic spectra and shade in red the flux contribution from important species (see label at top-left in each panel; this contribution also includes the continuum flux). We show the flux $F_\nu$ (in mJy) versus wavelength (in microns) for better visibility (the infrared is typically a hundred times fainter than the optical) and adopting the \sn\ distance. In each row, we adjust the ordinate range in order to encompass the strongest line from the selected species.
\label{fig_species_s15p2_at_300d}
}
\end{figure*}

Figure~\ref{fig_species_s15p2_at_300d} illustrates more explicitly for model s15p2 at 300\,d the individual contributions of important species between H and Ni. Because of the large Fe abundance at solar metallicity, its strong line blanketing power, and the significant contribution of the abundant H-rich material in the ejecta, Fe\one\ and Fe\two\ contribute considerably to the escaping SN radiation, primarily in the form of a forest of lines up to 1-2\mic, and with just a few strong forbidden lines at longer wavelength. Being relatively isolated, these infrared lines carry robust information on the ejecta composition and dynamics. All other elements in this model contribute a few lines, more or less scattered throughout the optical and infrared. For example, Ca contributes mostly with lines of Ca\two\ in the optical, whereas Ni contributes both Ni\one\ and Ni\two\ lines throughout the optical and infrared.

\subsection{Profile morphology of a sample of optical and infrared lines}

\begin{figure*}
\centering
\includegraphics[width=0.8\hsize]{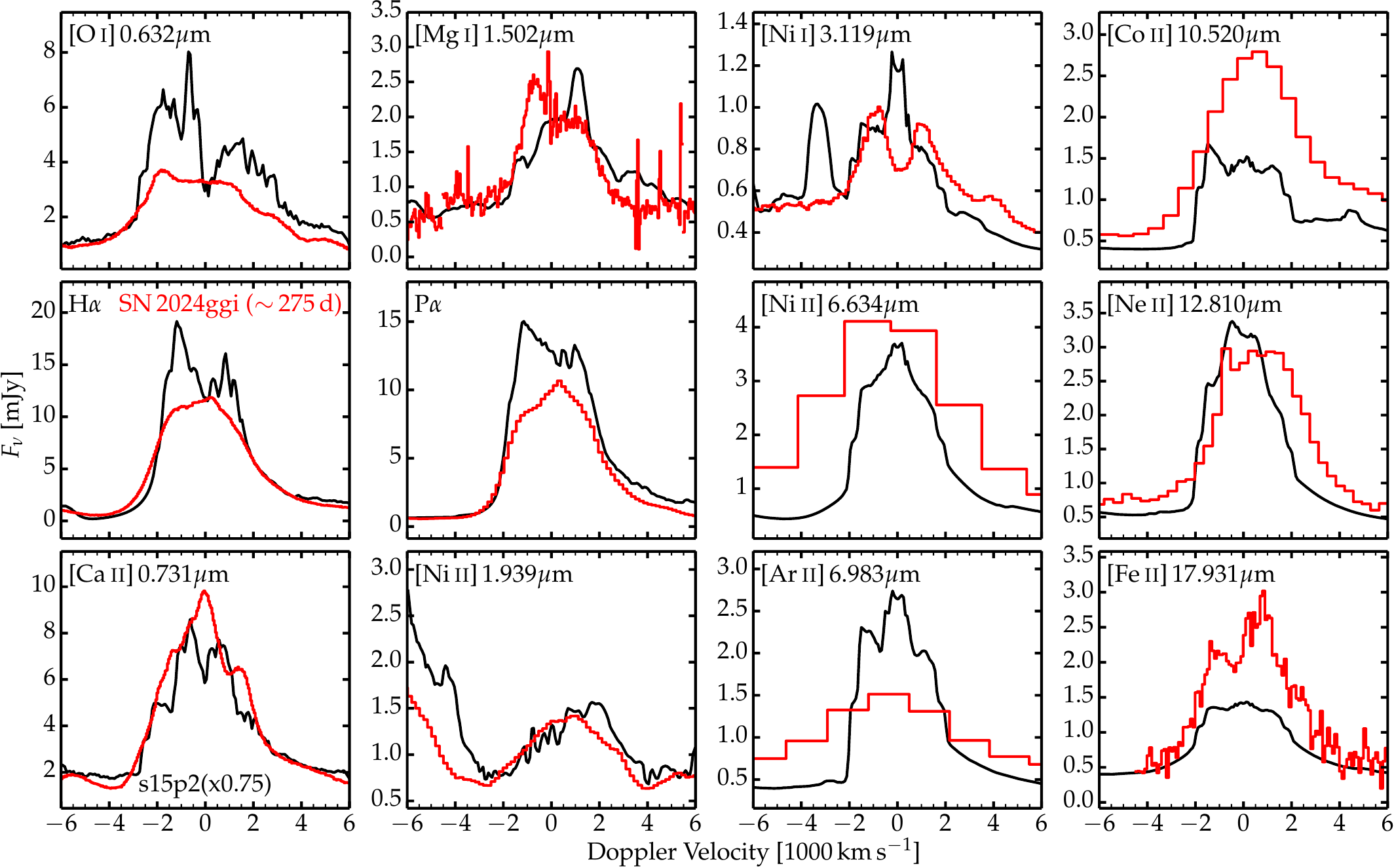}
\caption{Same as Fig.~\ref{fig_mod_vs_obs_jan25} but comparing the flux in individual emission lines versus Doppler velocity for the first epoch of observations of \sn\ with the s15p2 model at 275\,d. The $x$-axis origin corresponds to the line rest wavelength but because of line overlap (in the model) or asymmetry (only in the observations since the model is 1D), the emission may be skewed. The fluxes are shown as $F_\nu$, which yields a comparable flux level of a few mJy for all strong lines across the optical and infrared at the SN distance. Lines are ordered from top to bottom and left to right in order of increasing wavelength.
\label{fig_mod_vs_obs_indiv_lines_epoch1}
}
\end{figure*}

\begin{figure*}
\centering
\includegraphics[width=0.8\hsize]{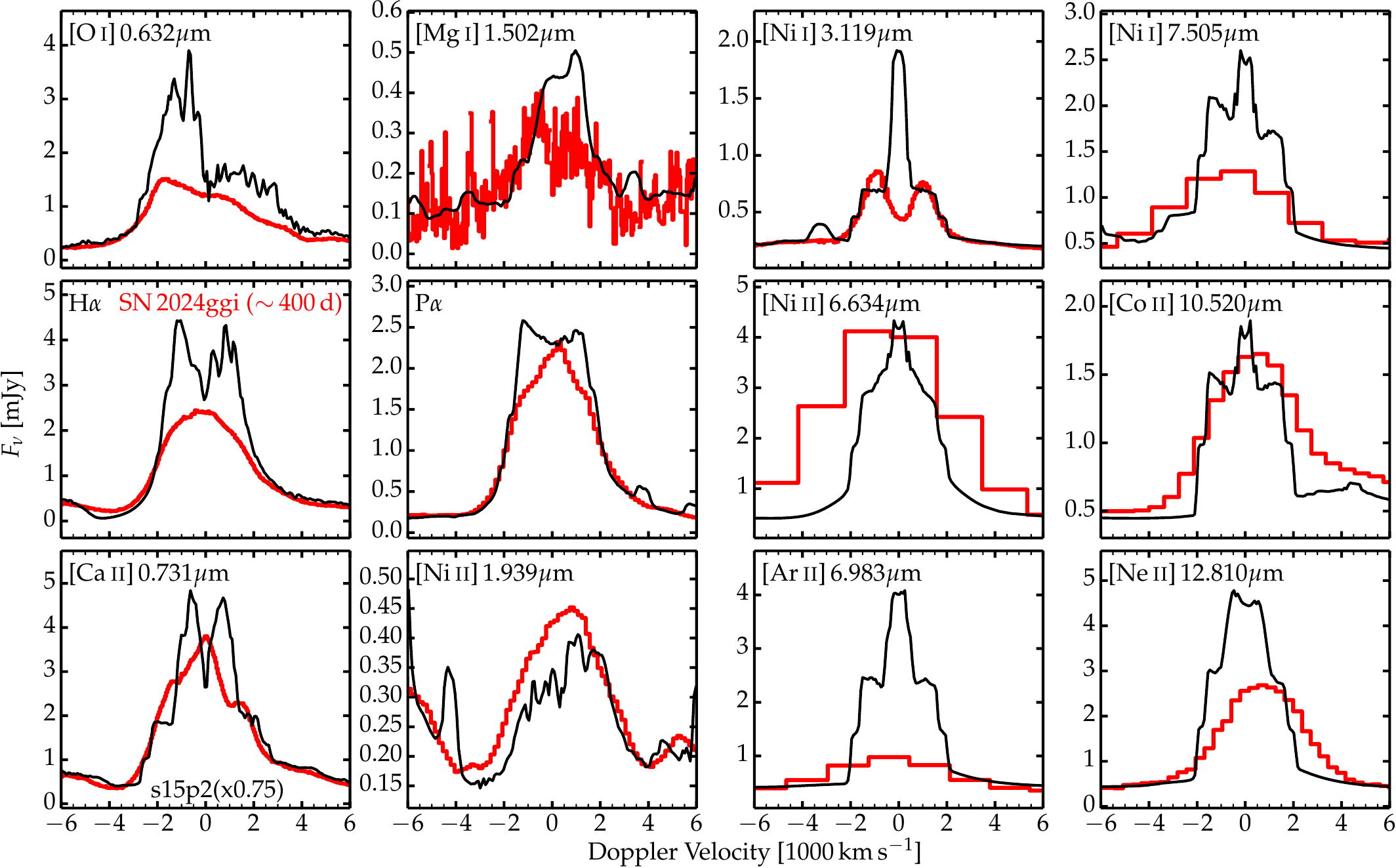}
\caption{Counterpart to Fig.~\ref{fig_mod_vs_obs_indiv_lines_epoch1} but for the second epoch of observations of \sn\ and the model at 400\,d, as shown versus wavelength in Fig.~\ref{fig_mod_vs_obs_may25}.
\label{fig_mod_vs_obs_indiv_lines_epoch2}
}
\end{figure*}

Figures~\ref{fig_mod_vs_obs_indiv_lines_epoch1} and \ref{fig_mod_vs_obs_indiv_lines_epoch2} illustrate the profile morphology of a variety of lines from H, O, Ne, Mg, Ar, Ca, Ni, and Co for both epochs of observations of \sn\ and for model s15p2 at 275\,d and 400\,d (the setup is the same as in Figs.~\ref{fig_mod_vs_obs_jan25} and \ref{fig_mod_vs_obs_may25}). Here, only a distance scaling is applied to the model (with an additional factor of 0.75), indicating a satisfactory match to the observations across the whole wavelength range as well as for most lines. An exception is, for example, the observed double-peak emission of [Ni\one]\,3.119\mic\ (see also discussion in \citealt{hueichapan_24ggi_25}) whereas the model yields an emission with a central peak. However, this unlikely reflects the complete distribution of Ni since the \nkiinir\ line has a different profile, this time well matched by the model. The different morphology of [Ni\one] and [Ni\two] lines indicates the presence of Ni-rich regions of different ionization, likely related to differences in density.

\section{Observational datasets}
\label{sect_obs}

We obtained two epochs of longslit spectroscopy of \sn\ on 26 January and 1 June 2025 and with the Keck Observatory Low Resolution Imaging Spectrometer (LRIS) \citep{oke_lris_95}. We used the 400/3400 grating for blue exposures and the 400/8500 grating for the red exposures, both with a 1.0'' slit mask. For all of these spectroscopic observations (see summary in Table~\ref{tab:spec_all}), standard CCD processing and spectrum extraction were accomplished with LPipe \citep{perley_lris_19}. The near-infrared spectra were obtained with the $R$\,$\sim$\,2700 Near-Infrared Echelle Spectrograph (NIRES; \citealt{wilson_spectro_04}) located on Keck-II on UT 15 January and 15 May 2025. For both observations we exposed for 1200\,s total divided into $4$\,$\times$\,300\,s individual exposures, dithering along the slit to allow for better sky subtraction. The data were reduced using a custom version of the IDL based reduction package Spextool \citep{cushing_spex_04} modified for use with NIRES. We used xtellcor \citep{vacca_nir_03} to correct for telluric features in our spectrum using an A0 standard star observed close in airmass and time to our target. For the spectrum obtained on 12 January 2025 we used HIP54427 and on 15 May 2025 we used HIP56308 for the telluric correction.

\begin{figure*}
\centering
\includegraphics[width=\hsize]{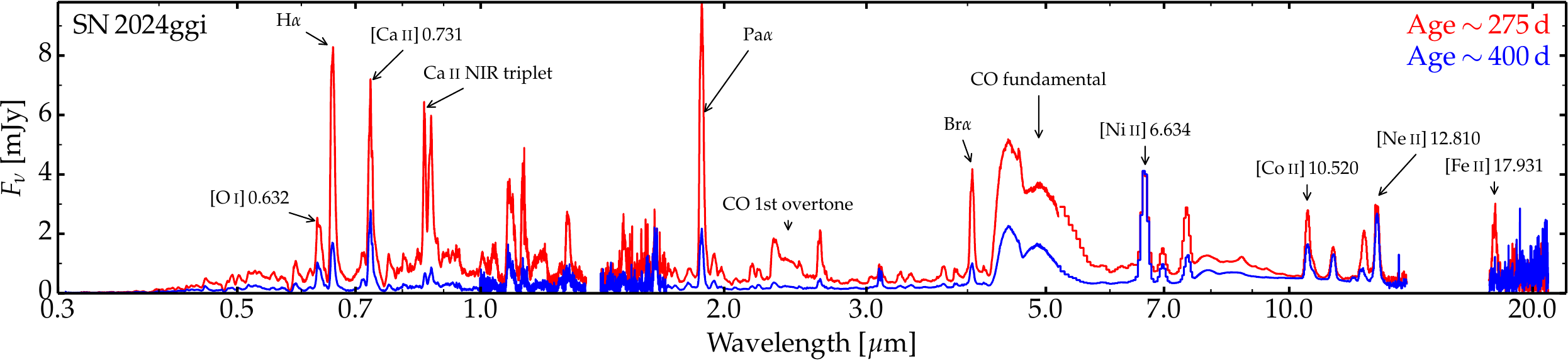}
\caption{Keck-LRIS optical, Keck-NIRES near-infrared, and JWST infrared spectroscopic observations of \sn\ obtained at about 275 and 400\,d after explosion. The data have been corrected for redshift.
\label{fig_obs_only}
}
\end{figure*}

We observed \sn\ on 1 January 2025 and 5 May 2025 corresponding to epochs of about +266 and +390\,d with the MIRI-LRS instrument on JWST that covers 5--14\mic\ with a resolving power ($R=\lambda/\Delta\lambda$) of $\sim$\,40--160 (programme Id\,\#\,6678, PI Kotak). The target was nodded along the slit to facilitate the removal of background and other detector systematics. The instrument configurations are listed in Table~\ref{tab:spec_jwst};  we used three integrations of 50 and 100 groups, respectively for the first and second epochs. In order to focus on particular spectral lines of interest, we also obtained medium resolution MRS spectra at the same epochs, using the SHORT grating settings for each of the 4 channels covering 4.90$-$5.74, 7.51$-$8.77, 11.55$-$13.47, and 17.70$-$20.95\mic\ with $R\sim3500-1700$ across this range. Given the brightness of the target, we proceeded with the Stage 3 products generated by v1.17.1 of the JWST pipeline \citep{bushouse_jwst_24} using the jwst\_1321.pmap Calibration Reference Data System (CRDS). However, we performed crosschecks starting with the uncalibrated frames and using v1.18.0 with jwst\_1364.pmap, and found the differences to be minimal. The resulting spectra, combined with the ground-based optical and near-infrared data described above, are shown in Fig. \ref{fig_obs_only}. The dataset used here was augmented with near contemporaneous NIRSPEC observations (programme Id\,\#\,6716, PI Ashall).

\begin{table*}
  \caption{Optical and near-infrared spectroscopic observations of \sn.}
  \label{tab:spec_all}
  \begin{center}
  \begin{tabular}{c@{\hspace{4mm}}c@{\hspace{4mm}}c@{\hspace{4mm}}c@{\hspace{4mm}}c@{\hspace{4mm}}c@{\hspace{4mm}}}
\hline
      UT Date & MJD & Phase$^a$ & Telescope & Instrument & Wavelength range [\AA]  \\
\hline
2025-01-12T13:52:18 & 60687.58 & 277.0 & Keck & NIRES & 9648-24664 \\
2025-01-26T13:18:43 & 60701.55 & 291.0 & Keck & LRIS & 3087-10280 \\
\hline
2025-05-15T06:10:47 & 60810.26 & 399.7 & Keck & NIRES & 9653-24668 \\
2025-06-01T06:03:38 & 60827.25 & 416.7 & Keck & LRIS & 3076-10276 \\
\hline
\end{tabular}
\end{center}
{\bf Notes:} $^a$: With respect to the time of first light ($\mathrm{MJD} = 60410.56$)
\end{table*}

\begin{table*}
\renewcommand{\arraystretch}{1.2}
\setlength\tabcolsep{0.1cm}
\fontsize{10}{11}\selectfont
\begin{center}
\caption{MIRI observations of \sn.}
\label{tab:spec_jwst}
\begin{tabular}{ccrccccc} \\
\hline
      UT date &
    MJD &
    Epoch$^{a}$ &
    Configuration &
    Exposure Time  \\
       &     &   (days)  &         &  (s) \\
\hline
2024 Jan. 22 & 60697.13 & 286.6 & NIRSPEC$^b$               &  192  \\
2024 Dec. 31 & 60675.90 & 265.3 & MRS SHORT, LONG (SHORTA)  & 2198  \\
2025 Jan. 01 & 60676.26 & 265.7 & LRS P750L                 &  844  \\
\hline
2025 May  02 & 60797.29 & 386.7 & NIRSPEC$^b$               &  472  \\
2025 May  05 & 60800.33 & 389.8 & LRS P750L                 & 1676  \\
2025 May  05 & 60800.36 & 389.8 & MRS SHORT, LONG (SHORTA)  & 2102  \\
\hline
\end{tabular}
\end{center}
{\bf Notes:} $^{a}$ With respect to the time of first light ($\mathrm{MJD} = 60410.56$). $^b$: From JWST programme Id.\,\#\ 6716.
\end{table*}

\end{document}